\documentclass[3p]{elsarticle}
\usepackage{bbm}
\usepackage{graphics}
\usepackage{graphicx}
\usepackage{verbatim}
\usepackage{subcaption}
\usepackage{amsmath, amssymb}
\usepackage{epsfig}
\usepackage[colorlinks]{hyperref}
\usepackage{soul}
\usepackage[utf8]{inputenc}
\usepackage{comment}
\usepackage{tikz}
\usetikzlibrary{
  arrows.meta,
  decorations.pathmorphing,
  positioning
}

\usepackage{caption}
\usepackage{lineno,hyperref}
\modulolinenumbers[5]

\journal{Journal of Applied Mechanics}

\begin{document}

\begin{frontmatter}

\title{On the Statistical Mechanics of Active Membranes: Some Selected Results} 

\author{Sreekanth Ramesh\fnref{1}}
\author{Prashant K. Purohit\corref{cor1}\fnref{2}}
\ead{purohit@seas.upenn.edu}
\author{Yashashree Kulkarni\corref{cor1}\fnref{1}}
\ead{ykulkarni@uh.edu}
\affiliation[1]{organization={Department of Mechanical and Aerospace Engineering, University of Houston},
             city={Houston},
          postcode={77204},
             state={TX},
             country={USA}}
\affiliation[2]{organization={Department of Mechanical Engineering and Applied Mechanics, University of Pennsylvania},
             city={Philadelphia},
             postcode={19104},
             state={PA},
             country={USA}}
\cortext[cor1]{Corresponding author}

\begin{abstract}
Biological membranes and vesicles play a central role in living systems, forming dynamic interfaces that regulate cellular organization and function. Classical descriptions of membrane mechanics that are rooted in equilibrium statistical mechanics and linear elasticity have yielded deep insights into membrane morphology and the role of thermal fluctuations on cellular function. However, real biological membranes operate far from equilibrium, continuously driven by active processes powered by energy-consuming proteins. In this work, we employ a non-equilibrium statistical mechanics framework to model active membranes and derive analytical expressions for four fundamental properties that characterize their mechanical behavior: (a) the tension–area relation, (b) the mean square amplitude of fluctuations, (c) correlation of normal vectors, and (d) the persistence length. These results collectively highlight the utility of fluctuation spectra as a starting point for elucidating membrane mechanics in both passive and active settings. Moreover, these results provide a theoretical basis for analyzing and interpreting fluctuation-based assays of active membrane behavior. 
\end{abstract}


\end{frontmatter}



\section{Introduction}

Biological membranes and vesicles are indispensable components of life, providing the essential interfaces that define cellular boundaries and mediate critical physiological processes. For decades, the mechanics of these systems has been elegantly described through the lens of equilibrium statistical mechanics and continuum mechanics, primarily building upon the foundation laid by the works of Helfrich, Canham and Evans \cite{Helfrich_1973, Canham, evans1974bending}. These studies have not only provided phenomenal insights into the mechanical behavior of biomembranes \cite{safran, nelson, boal2012mechanics, steigmann-book, Kaisar2025AMR, steigmann-a, Rangamani2014Variable, purohit_2016, dayal_instability_2016, steigmann2021} but also elucidated the role of thermal fluctuations of these entities in explaining diverse biological phenomena such as membrane inclusions \cite{purohit_inclusions_2018, agrawal_2016}, vesicle size distributions \cite{helfrich_1986, ahmadpoor_sharma_2016}, membrane steric interactions \cite{helfrich_1978, Gompper_1989, freund_2012, sharma_entropic_2014, deshpande_2019, hassan2025}, role of cytoskeleton \cite{seifert_2019, steigmann2020} and electromechanical coupling in membranes \cite{purohit_flexo_2013, liping_2013, grasinger2021, torbati_rmp_2022}. 

While these studies have provided profound insights into membrane fluctuations and morphology, they essentially describe a passive or ``nonliving'' state. However, in reality, biological membranes are now known to be fundamentally active or ``alive'' constantly driven away from equilibrium by active proteins that harness energy from sources such as Adenosine Triphosphate (ATP) hydrolysis or light. Starting with the pioneering studies by Prost, Ramaswamy and others \cite{prost_1996, rama_prost_1999, rama_2001}, there is a growing body of research elucidating the mechanical response of active matter and its vital role in various physiological processes and phenomena \cite{gov_2004, gov_2006, seifert_2012, purohit_active_2020, takatori_sahu_2020, kulkarni2023fluctuations, ramesh2024statistical, mathew2025pnas, Kaisar2025}.

Although the field of active matter is vast, encompassing active Brownian particles, active filaments, active membranes, active gels and so on \cite{rama_2010, turlier_betz_2019, marchetti2013review, needleman2017review}, here we focus on active biological membranes. In this paper, we apply the non-equilibrium statistical mechanics framework developed in several earlier works to derive a collection of theoretical results for membranes exhibiting thermal and active fluctuations. Specifically, we present the derivation for four fundamental properties that characterize the mechanical behavior of active membranes: (a) tension-area relation, (b) mean square amplitude of fluctuations, (c) correlation between normal vectors, and (d) persistence length for active membranes. The purpose of this analysis is two-fold. First, it illustrates the value of deriving the fluctuation spectra for a fluctuating membrane as it serves as the starting point for extracting different insights into the mechanical response of passive and active membranes. Second, these results are important in themselves as these are quantities that can be measured in experiments, and hence our analytical expressions can serve as a theoretical tool for interpreting measurements for active membranes. Furthermore, these results can also be used to identify signatures of activity in biological membranes. 

\section{A continuum theory for active membranes}\label{Math_prelim_membrane_subsection}
We consider the membrane to be an elastic sheet that is resistant to change in area,
but can bend easily by deforming out-of-plane~\cite{steigmann-book, Phillips2013}. Mathematically, we consider it as a smooth, orientable surface $\mathcal{S}$ embedded in 3-dimensional space with $\mathbf{n}$ being a unit normal vector field on the surface. This implies that each point on $\mathcal{S}$ will have an  unique normal $\mathbf{n}$ associated with it. Biological membranes are  primarily regarded as fluid membranes that typically possess only bending elastic energy. Accommodating the areal constraint, the total potential energy of the membrane (active or passive) is defined as \cite{steigmann-book, biria_2013}
\begin{equation}\label{membrane-energy}
\mathcal{F} = \int_\mathcal{S} (\psi (H,K) + \sigma)\,  d\mathcal{S}
\end{equation}
where $\psi(H,K)$ is the elastic energy density and is a function of the mean curvature $H$, and the Gaussian curvature $K$. Mathematically, $\sigma$ is the Lagrange multiplier associated with the areal constraint. Physically, $\sigma$ is the surface tension in the membrane that describes the energy cost for change in membrane area due to deformation. Most studies based on linearized
curvature elasticity assume the elastic energy density to be quadratic and expressed using the renowned Helfrich–Canham–Evans form \cite{Helfrich_1973, Canham, evans1974bending},
\begin{equation}\label{Helfrich-hamiltonian}
\psi (H,K) = \frac{1}{2} \kappa (H-H_0)^2 + \bar{\kappa} (K-K_0),
\end{equation}
where $\kappa$ and $\bar{\kappa}$ represent the bending moduli for the respective changes in curvature. Here, $H_0$ and $K_0$ represent the corresponding spontaneous curvature that the membrane has in the absence of applied forces and moments. In this work, we will also use Eq.~\eqref{Helfrich-hamiltonian} for the membrane elastic energy.  

\begin{figure}[b!]
  \centering
  \includegraphics[width=0.7\textwidth]{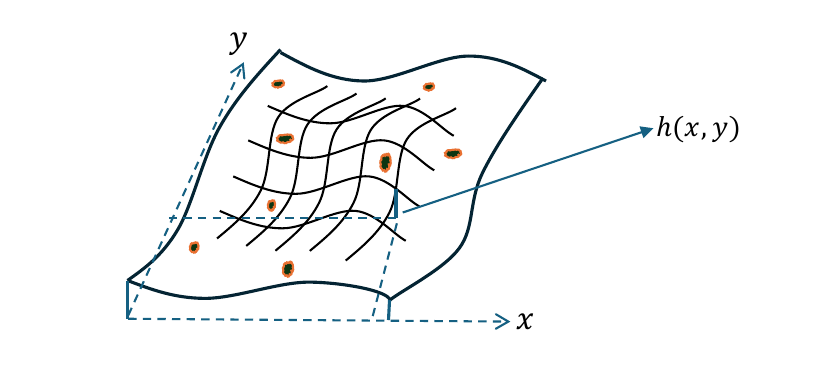}
  \caption{A schematic of a fluctuating active membrane showing the out-of-plane fluctuations $h(x,y)$. The red objects embedded on the membrane are molecular motors that consume ATP and exert random forces on the membrane in addition to the buffetting by Brownian forces.}
\end{figure}

The commonly used approach for deriving the governing equation for an active membrane is to solve the Stokes equation in the surrounding fluid, and apply displacement and traction boundary conditions at the interface of the fluid and membrane \cite{gov_2006, kulkarni2023fluctuations, turlier_betz_2019, seifert_1997}. This is a reasonable assumption since the cellular fluid has a very low Reynolds number and hence inertial forces are negligible. This yields an overdamped Langevin equation for an active membrane. We propose an alternative approach here based on a variational formulation to derive the shape equation for the membrane and then incorporating the dissipative effects from the embedding fluid. 

To this end, we assume the membrane to be fluctuating in a Stokes fluid and consider an instantaneous membrane configuration with $\mathbf{x}$ denoting the position vector of a generic point on its surface. Let $\mathbf{x}$ be parametrized by a set of local coordinates, say $s_1$ and $s_2$. 
The action is defined as the time integral of the negative of the Hamiltonian \cite{lanczos1986}, which in the absence of a kinetic energy term becomes 
\begin{equation}
    A \equiv \int_{t_1}^{t_2} \left(-\int_\mathcal{S} (\psi (H,K) + \sigma)\,  d\mathcal{S}\right) dt.
\end{equation}
We now perturb the system such that we have a  new configuration $\mathbf{x}_\epsilon$ as \cite{biria_2013}
\begin{equation}
    \mathbf{x}_\epsilon = \mathbf{x}  + \epsilon \mathbf{u}\,,
\end{equation}
where $\epsilon$ is a small parameter and $\mathbf{u}$ is a variation which we decompose as 
\begin{equation}
   \mathbf{u} = \mathbf{u}_t + U \mathbf{n}\,. 
\end{equation}
Here, $\mathbf{u}_t = \mathbf{P}\mathbf{u}$ is the tangential component of $\mathbf{u}$ and $U = \mathbf{u} \cdot \mathbf{n}$ is the scalar normal component of $\mathbf{u}$ with $\mathbf{P}$ being the surface projection tensor and expressed as
\begin{equation}
    \mathbf{P} = \mathbf{1} - \mathbf{n} \otimes \mathbf{n}\,.
\end{equation}
Thus, the variation of the action is given by \cite{biria_2013}
\begin{multline}
    \delta A = \int_{t_1}^{t_2} \biggl(-\int_\mathcal{S} \Bigg[ 
{\psi}_H (2H^2 - K) + \frac{1}{2} \Delta_S {\psi}_H + 2 {\psi}_K HK + 2 \Delta_S ({\psi}_K H) \\
- \mathrm{div}_S (\mathbf{L} \nabla_S {\psi}_K) - 2 (\nabla_S H) \cdot (\nabla_S {\psi}_K) - 2 {\psi}_K \Delta_S H - 2H (\psi + \sigma)\Bigg] U \biggr) \,dt\,
\label{action_variation}
\end{multline}
We refer the reader to \cite{biria_2013} for a detailed derivation. Here, we have ignored the edge terms since we restrict ourselves to membranes with periodic boundary conditions. 
 In Eq.~\eqref{action_variation}, $\psi_H=\frac{\partial\psi}{\partial H}$, $\psi_K=\frac{\partial\psi}{\partial K}$ and the curvature tensor $\mathbf{L}$, mean curvature and Gaussian curvature are respectively defined as
\begin{equation}\label{curvaturetensor}
\mathbf{L} = - \mathrm{div}_S \mathbf{n}\,,\,\, H = \frac{1}{2} \mathrm{tr} \mathbf{L},\,\, K = \frac{1}{2} [(\mathrm{tr} \mathbf{L})^2 - \mathrm{tr} \mathbf{L}^2]
\end{equation}
The subscript $S$ attached to the $\nabla, \Delta$ and $\mathrm{div}$ operators in the above equations represent the surface gradient, Laplacian and divergence respectively, which simply means that they are the projections of their three-dimensional counterparts onto the surface $\mathcal{S}$.


Finally, the stationarity of the action, $\delta A = 0$, gives us the well-known shape equation for membranes \cite{steigmann-a, biria_2013, steigmann_2003}
\begin{multline}
    \int_\mathcal{S} \Bigg[ 
{\psi}_H (2H^2 - K) + \frac{1}{2} \Delta_S {\psi}_H + 2 {\psi}_K HK + 2 \Delta_S ({\psi}_K H) \\
- \mathrm{div}_S (\mathbf{L} \nabla_S {\psi}_K) - 2 (\nabla_S H) \cdot (\nabla_S {\psi}_K) - 2 {\psi}_K \Delta_S H - 2H (\psi+\sigma) \Bigg] U = 0.
\end{multline}
This is the governing equation for membranes in equilibrium. For the dynamic analysis of active membranes, we now append to this equation a frictional force term, a thermal noise term and an active noise term to obtain the over-damped Langevin equation, viz.,
\begin{multline} \label{Langevin_membrane_general}
 \Lambda \frac{\partial \mathbf{x}(s_1,s_2,t)}{\partial t} = \int_S \Bigg[ 
{\psi}_H (2H^2 - K) + \frac{1}{2} \Delta_S {\psi}_H + 2 {\psi}_K HK + 2 \Delta_S ({\psi}_K H) \\
- \mathrm{div}_S (\mathbf{L} \nabla_S {\psi}_K) - 2 (\nabla_S H) \cdot (\nabla_S {\psi}_K) - 2 {\psi}_K \Delta_S H - 2H (\psi+\sigma) \Bigg]U + \xi^{th}(s_1,s_2,t) + \xi^{a}(s_1,s_2,t),
\end{multline}
where ${\xi^{th}(s_1,s_2,t)}$ is a thermal noise having zero mean and correlation function 
\begin{equation}\label{thermal-noise-general}
   \langle \xi^{th}(s_1,s_2,t) \xi^{th}({s_1}',{s_2}',t')\rangle = 2\Lambda k_B T \delta(s_1-{s_1}')\delta(s_2-{s_2}')\delta(t-{t}')
\end{equation}
in order to satisfy the fluctuation dissipation theorem~\cite{pathria}. Here $\Lambda$ is the viscosity of the embedding fluid and $k_{B}T$ is the unit of thermal energy at absolute temperature $T$ and $\delta$ is the dirac delta function. Unlike the noise due to thermal fluctuations, the active noise $\xi^{a}(s_1,s_2,t)$ can be correlated in space and time, having different forms for the correlation function. Based on experimental evidence, the active noise is conventionally considered to be uncorrelated in space with an exponentially decaying correlation in time. With $\Gamma^a$ denoting the strength of the force from active proteins and $\tau_a$ being the characteristic time for the proteins switching between `ON' and `OFF' states, we can write the correlation function for the active noise as follows\cite{rama_2001, gov_2004,seifert_2012, turlier_betz_2019}:  
\begin{equation}\label{active-noise-general}
    \langle \xi^{a}(s_1,s_2,t) \xi^{a}({s_1}',{s_2}',t')\rangle =\Gamma^\mathrm{a} \,\delta(s_1-{s_1}')\delta(s_2-{s_2}') e^{-(t -t^\prime)/\tau^\mathrm{a}}.
\end{equation}
Equation \eqref{Langevin_membrane_general} together with Eq.~\eqref{thermal-noise-general} and Eq.~\eqref{active-noise-general} yields the over-damped Langevin equation that governs the non-equilibrium statistical mechanics of an active membrane. 

t\section{Fluctuations of a quasi-planar active membrane} \label{planar_membranes}

For the sake of simplicity, we now consider a quasi-planar membrane of contour area $A$ that fluctuates around a mean flat configuration and assume periodic boundary conditions. A convenient description for such a system is the Monge parametrization, where the out-of-plane deformation is described by the height $h(\mathbf{r})$, with $\mathbf{r} = (x,y)^T$ being the position vector in the flat undeformed configuration \cite{Deserno-primer}. 

The unit normal of the membrane is expressed as 
\begin{equation}\label{normal_vector}
    \mathbf{n} = \frac{1}{\sqrt{1 + |\nabla h(\mathbf{r})|^2}} ( -\nabla h(\mathbf{r}) ,\\ 1 )
\end{equation}
from which we can write the mean and Gaussian curvature as \cite{torbati_rmp_2022, deserno_2015}
\begin{equation}\label{Mean_and_gaussian_curvature}
    H = \nabla \cdot \left( \frac{\nabla h(\mathbf{r})}{\sqrt{1 + |\nabla h(\mathbf{r})|^2}} \right), \quad
K = \frac{\det(\nabla \nabla h(\mathbf{r}))}{\left(1 + |\nabla h(\mathbf{r})|^2\right)^2}.
\end{equation}
In the above expressions, $\nabla$ represents the gradient with respect to the $x$ and $y$ coordinate system.

Under small gradient approximation, $\nabla h << 1$, we can assume $H \approx \Delta h(\mathbf{r})$. Also, since we consider membranes without any edges or holes, the contribution of the Gaussian curvature to the potential energy vanishes by virtue of the Gauss-Bonnet theorem \cite{boal2012mechanics}. Substituting this,  along with the areal element $d\mathcal{S} = \sqrt{1 + |\nabla h(\mathbf{r})|^2}\, dx\, dy$, the total potential energy in Eq.~\eqref{membrane-energy} can be written as,
\begin{equation}\label{Potential_energy_flat}
    \mathcal{F} = \frac{1}{2} \int_{S_0} dx\,dy \left[ \kappa (\Delta h)^2 + \sigma |\nabla h|^2 \right]
\end{equation}
with $S_0$ being the flat parametric domain and $\sigma$ the applied tension. Since the energy is quadratic in $\Delta h$ and $\nabla h$ we will exploit a Fourier series expansion of the height function $h$: 
\begin{equation}\label{h_fourier}
h(\mathbf{r}) = \sum_{\mathbf{q}} h_\mathbf{q} e^{i \mathbf{q}\cdot \mathbf{r}}\,, 
\end{equation}
where $\mathbf{q}$ is a wave vector.\footnote{For a membrane having area $A = L \times L$, %
\[
q := |\mathbf{q}| \in [q_{\min}, q_{\max}], \quad \text{i.e.}
\]
\[
\mathcal{K} =
\left\{
\mathbf{q} :
\mathbf{q} = \frac{2\pi}{L}(v_x, v_y),\;
v_x, v_y \in \mathbb{Z},\;
|\mathbf{q}| \in [q_{\min}, q_{\max}]
\right\}.
\]
}
This allows us to write the potential energy in terms of the modes $h_\mathbf{q}$ of the system, giving us a quadratic form for the Hamiltonian \cite{safran, boal2012mechanics, Deserno-primer}: 
\begin{equation}\label{Energy_fourier}
\mathcal{F} = \frac{A}{2} \sum_\mathbf{q}|h_\mathbf{q}|^2 \,\left[\kappa q^4 + \sigma q^2\right]\,.
\end{equation}
To get the dynamical equation for a quasi-planar membrane, we simply substitute equations Eq.~\eqref{normal_vector} and Eq.~\eqref{Mean_and_gaussian_curvature} into Eq.~\eqref{Langevin_membrane_general}. Expanding the height $h(\mathbf{r})$ in terms of the Fourier modes, we can write the over-damped Langevin equation for each mode of a quasi-planar active membrane as 
\begin{equation}\label{Langevin-monge_active}
\frac{\partial h_\mathbf{q} (t)}{\partial t} = \frac{A}{\Lambda} \left[ - (\kappa q^4 + \sigma q^2) h_\mathbf{q} (t) + \xi^\mathrm{th}_\mathbf{q} (t) + \xi^\mathrm{a}_\mathbf{q} (t)\right] = - \omega_\mathbf{q} h_\mathbf{q} (t) + \frac{A}{\Lambda}[ \xi^\mathrm{th}_\mathbf{q} (t) + \xi^\mathrm{a}_\mathbf{q} (t)], 
\end{equation} 
where $\omega_q = \frac{A}{\Lambda}(\kappa q^4 + \sigma q^2)$ is a frequency corresponding to wave number $q$ and the correlation functions for the noise terms reduce to 
\begin{equation}
     \langle \xi^\mathrm{th}_\mathbf{q} (t) \xi^\mathrm{th}_{\mathbf{-q}} (t^\prime) \rangle = \frac{2\Lambda k_B T}{A} \delta(s_1 -s_2).
 \end{equation}
and
\begin{equation} \label{active-noise}
\langle \xi^\mathrm{a}_\mathbf{q} (t) \xi^\mathrm{a}_{\mathbf{q}^\prime} (t^\prime) \rangle = \frac{\Gamma^\mathrm{a}_q}{A} \,\delta(\mathbf{q} - \mathbf{q}^\prime) e^{-|t -t^\prime|/\tau^\mathrm{a}}.
\end{equation} 
We note that the subscript $q$ in $\Gamma^\mathrm{a}_q$ refers to the fact that the activity strength may depend on the wave number $q$. In prior studies, there are two common ways of modeling activity using either direct force (where proteins randomly exert direct force on the membrane) or curvature force (where active proteins exert force by modifying the curvature) \cite{gov_2006, kulkarni2023fluctuations}. In the former case, $\Gamma^\mathrm{a}_q$ does not depend on $q$ but in the latter case it does. In subsequent calculations, we will work with direct force only. 

From Eq.~\eqref{Langevin-monge_active}, we can get the steady state fluctuation spectra as, 
\begin{equation}\label{Fluctuation_spectra_active_beg}
    \langle |h_{\mathbf{q}}|^2 \rangle = \frac{A}{\Lambda^2} \lim_{t \to \infty} \int_0^t ds_1 \int_0^t ds_2 \, \{ \langle \xi^\mathrm{th}_\mathbf{q}(s_1) \xi^\mathrm{th}_\mathbf{-q}(s_2) \rangle + \langle \xi^\mathrm{a}_\mathbf{q}(s_1) \xi^\mathrm{a}_\mathbf{-q}(s_2) \rangle\}e^{\omega_{\mathbf{q}}(s_1 + s_2 - 2t)}.
 \end{equation} 
On the right hand side the integral for thermal noise is well known~\cite{boal2012mechanics}. Substituting the correlation for the active noise into Eq.~\eqref{Fluctuation_spectra_active_beg}, we have
\begin{eqnarray}\label{Fluctuation_spectra_active}
   \langle |h_{\mathbf{q}}|^2 \rangle &=& \frac{k_B T}{A(\kappa q^4 + \sigma q^2)} + \frac{1}{\Lambda^2} \lim_{t \to \infty} \int_0^t ds_1 \int_0^t ds_2 \, \Gamma^\mathrm{a}_q \, e^{-|s_1 -s_2|/\tau^\mathrm{a}} e^{\omega_{\mathbf{q}}(s_1 + s_2 - 2t)}\\
   &=& \frac{k_B T}{A(\kappa q^4 + \sigma q^2)} + \frac{ \Gamma_q^a}{\Lambda^2\omega_q}\frac{1}{(\omega_q + 1/\tau_a)},
\end{eqnarray}
 This result for the fluctuation spectra is consistent with previously derived results \cite{gov_2006, kulkarni2023fluctuations}. It reveals how activity influences fluctuations and provides a means to estimate the properties of active membranes through experimentally observed fluctuations. In subsequent sections, we will use this result to derive analytical expressions for various quantities characterizing active membranes. 

\section{Tension-Area relation}
We can now derive a tension-area relation for membranes. The area we refer to here is the projected area of the membrane. The tension-area relation accounts for the response of the membrane to applied tension $\sigma$, by calculating the area stored in the undulations due to fluctuations. The contour area $A$ of a given configuration is given by \cite{boal2012mechanics}
\begin{equation}
A \cong \int d\mathbf{r} + (1/2) \int (h_x^2 + h_y^2)\, d\mathbf{r},
\end{equation}
where, $(1/2)\int (h_x^2 + h_y^2)d\mathbf{r}$ is the amount by which the projected area $\int d\mathbf{r}$ is reduced below its contour area $A$ by thermal fluctuations. This reduction can be be obtained by averaging over all configurations of the ensemble:
\begin{equation}\label{contour_area}
\langle \int (h_x^2 + h_y^2) d\mathbf{r} \rangle = (A^2 / 4\pi^2) \int q^2 d\mathbf{q} \, \langle |h_{\mathbf{\mathbf{q}}}|^2 \rangle.
\end{equation}
Substituting the expression for the fluctuation spectrum in Eq.~\eqref{contour_area}, setting $\Gamma^a_q = 0$ (for now), and observing that the order of integration and ensemble averaging can be switched (because both are sums), we have
\begin{equation}\label{Tension_area_relation}
A_{\text{red}} = (1/2) \int \langle h_x^2 + h_y^2\rangle d\mathbf{r} = A (k_B T / 8\pi^2) \int d\mathbf{q} / (\kappa q^2 + \sigma),
\end{equation}
which, on simplifying further, reduces to 
\begin{equation}
A_{\text{red}}(\sigma)/A = (k_B T / 8\pi) \int dq^2 / (\kappa q^2 + \sigma).
\end{equation}
The limits of the above integral for the wave vector are determined by the largest and smallest length scales of the system. For a membrane having area $A$ and lipid headgroup size $b$ the integral is evaluated from $q_{min} = \pi / \sqrt{A} $ to $q_{max} = \pi /b$, giving us the following expression,
\begin{equation} \label{eq:classtenar}
\frac{A_{\mathrm{red}}(\sigma)}{A}
= \frac{k_B T}{8\pi \kappa}
\ln\!\left(
\frac{\pi^2/b^2 + \sigma/\kappa}{\pi^2/A + \sigma/\kappa}
\right).
\end{equation}
\begin{figure}[t!]
  \centering
  \includegraphics[width=0.6\textwidth]{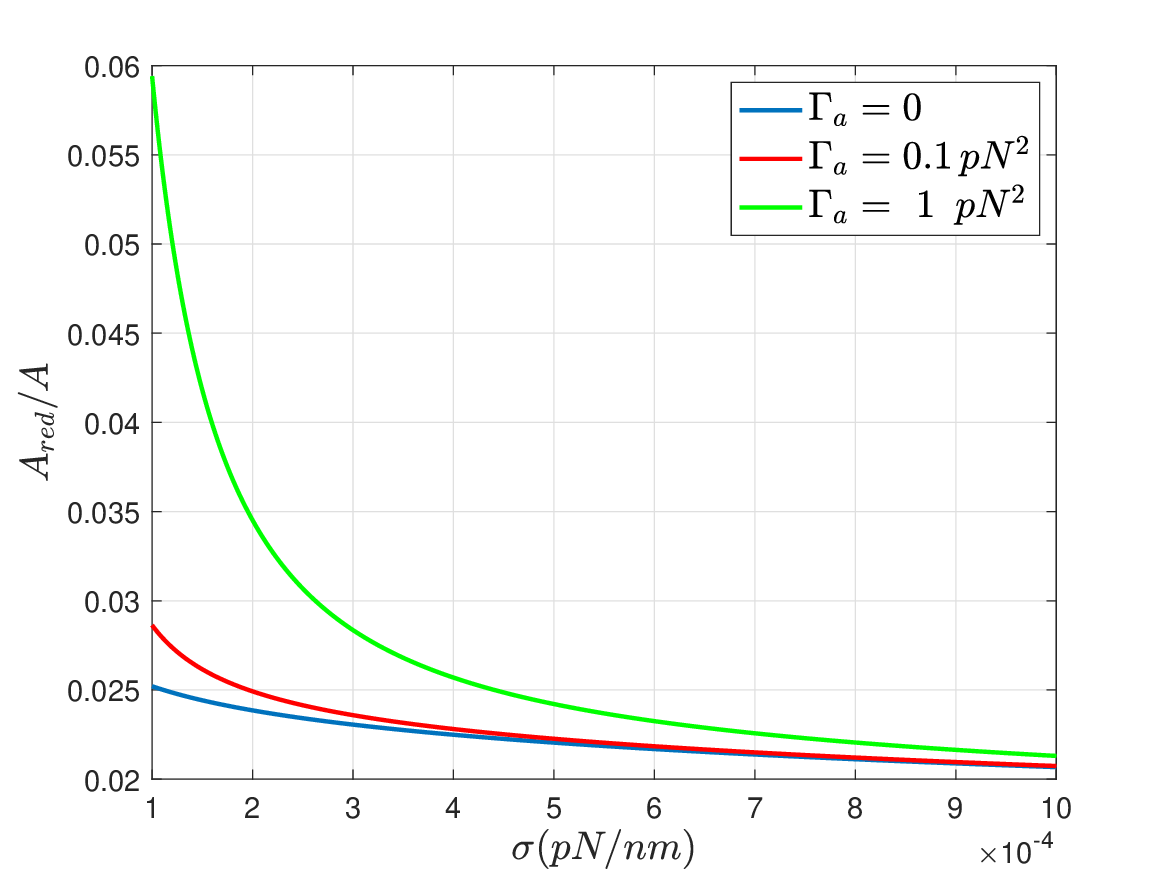}
  \caption{Reduction in projected area against tension for various levels of activity. Higher activity leads to more membrane area residing in shape fluctuations.}
  \label{Reduced_area}
\end{figure}

Eq. (\ref{eq:classtenar}) is the classic tension-area relation due to thermal fluctuations that can be found in biophysics text books \cite{boal2012mechanics, Phillips2013}. We can obtain the tension-area relation for an active membrane as follows by plugging in the fluctuation spectra in the presence of activity in Eq.~\eqref{contour_area}, viz.,
\begin{eqnarray}
 \int \langle h_x^2 + h_y^2 \rangle\, d\mathbf{r}
&=& \frac{A^2}{4\pi^2} \int_{q_{\min}}^{q_{\max}} q^2\, d\mathbf{q}\, 
\langle |h_{\mathbf{q}}|^2 \rangle \nonumber \\
&=& \frac{A^2}{4\pi^2} \int_{\pi/\sqrt{A}}^{\pi/b}  q^2\, d\mathbf{q}\, \left( \frac{k_B T}{A(\kappa q^4 + \sigma q^2)} + \frac{ \Gamma_q^a}{\Lambda^2\omega_q}\frac{1}{(\omega_q + 1/\tau_a)} \right) \nonumber \\
&=& \frac{A}{4\pi} \int_{\pi/\sqrt{A}}^{\pi/b} \, dq^2 \, \left( \frac{k_B T}{(\kappa q^2 + \sigma )} + \frac{ \Gamma_q^a}{\Lambda(\kappa q^2 + \sigma )}\frac{1}{(\omega_q + 1/\tau_a)} \right).
\end{eqnarray}
Therefore, from Eq.~\eqref{Tension_area_relation}, we arrive at
\begin{equation}
    \frac{A_{\text{red}}(\sigma)}{A} = \frac{1}{8\pi} \int_{\pi^2/A}^{\pi^2/b^2}  \, dq^2 \, \left( \frac{k_B T}{(\kappa q^2 + \sigma )} + \frac{ \Gamma_q^a}{\Lambda(\kappa q^2 + \sigma )}\frac{1}{(\omega_q + 1/\tau_a)} \right).
\end{equation}
We evaluate the above integral for $\tau_a \to \infty$ under the assumption that $\Gamma^a_q$ is independent of $q$ (as is the case for direct force) and is denoted by $\Gamma^a$. This yields
\begin{equation} \label{eq:fintA}
    \frac{A_{\text{red}}(\sigma)}{A} = 
\frac{k_B T}{8\pi\kappa} \,
\ln\!\frac{A\left(\pi^{2}\kappa + b^{2}\sigma\right)}
     {b^{2}\left(\pi^{2}\kappa + A\sigma\right)}
+ \frac{\Gamma^{a}}{8\pi A\sigma^{2}}\left[\pi^{2}\kappa
\left(\frac{1}{\pi^{2}\kappa + A\sigma}
-\frac{1}{\pi^{2}\kappa + b^{2}\sigma}\right)
-\ln\!\frac{\pi^{2}\kappa + b^{2}\sigma}{\pi^{2}\kappa + A\sigma}\right].
\end{equation}
The integral above can also be performed for a finite and non-zero $\tau_{a}$ resulting in analytical expressions (involving $\tan^{-1}$ and $\ln$ functions), but they are omitted here for brevity and also because key insights can be gleaned from Eq.~\eqref{eq:fintA}. In Eq.~\eqref{eq:fintA}, the first term on the right hand side is the contribution of thermal fluctuations and the second term arises from activity. For a tensionless membrane, with $\sigma = 0$, the expression for the reduction in area is further simplified to:
\begin{equation}
    \frac{A_{\text{red}}}{A} = \frac{1}{8\pi}\left[\frac{k_{B}T}{\kappa}\ln\frac{A}{b^{2}} + \frac{\Gamma^a}{2A\pi^4 \kappa^{2}}(A^2 - b^4)\right].
\end{equation}

Fig.~\ref{Reduced_area} shows curves of reduced area versus surface tension for various levels of activity calculated using Eq.~\eqref{eq:fintA}. The plot highlights two key insights. First, an increase in surface tension suppresses fluctuations, decreasing the difference $A_{\text{red}}$ between the contour area and the projected area, so the projected area increases as tension increases. Second, an increase in activity increases fluctuations and increases $A_{\text{red}}$, thereby decreasing projected area. Interestingly, $A_{\text{red}}$ depends linearly on both $k_{B}T$ and $\Gamma^{a}$. Therefore, it may not be straight-forward to distinguish the source of fluctuations in a live cell in which both Brownian (thermal) fluctuations and non-Brownian activity are taking place. 

\section{Mean square amplitude of thermal and active undulations}
Till now, we have studied the effects of fluctuations by using the expression for the mean-square amplitude of each of the Fourier modes. Rather than looking at individual modes, it is also useful to calculate the actual mean-square height fluctuations of membranes because these can be measured experimentally. From \cite{boal2012mechanics}, we have 
\begin{align}
\langle h^2 \rangle
&= \frac{A}{4\pi^2} \int d\mathbf{q}\,
\langle h(\mathbf{q}) h^*(\mathbf{q}) \rangle .
\end{align}
Substituting the fluctuation spectra for an active membrane we have, 
\begin{align} \label{eq:h2integ}
   \langle h^2 \rangle &= \frac{A}{4\pi^2} \int  d\mathbf{q}\, \left\{\frac{k_B T}{A(\kappa q^4 + \sigma q^2)} + \frac{ \Gamma_q^a}{\Lambda^2\omega_q}\frac{1}{(\omega_q + 1/\tau_a)} \right\}.
\end{align}
In Eq.~\eqref{eq:h2integ}, we re-write $\int d\mathbf{q}$ in terms of polar coordinates as the double integral $\int\int q dq d\theta$. Since the integrand is independent of $\theta$ we can integrate it out and the double integral reduces to an integral over $q$ alone, viz., $\int 2 \pi q\, dq$. Using this and substituting the expression for $\omega_q$, we calculate the integral by taking the limit of $\tau_a \to \infty$, and assuming again that $\Gamma^{a}_{q}$ is independent of $q$. This yields
\begin{align} \label{eq:finh2}
   \langle h^2 \rangle &= \frac{A}{2 \pi} \int_{\pi/\sqrt{A}}^{\pi/b} q\, dq\, \left\{\frac{k_B T}{A(\kappa q^4 + \sigma q^2)} + \frac{\Gamma^a}{A^2(\kappa q^4 + \sigma q^2)^2} \right\} \nonumber \\
   &= \frac{1}{2 \pi} \int_{\pi/\sqrt{A}}^{\pi/b} \, dq\, \left\{\frac{k_B T}{(\kappa q^3 + \sigma q)} + \frac{ \Gamma^a}{A(\kappa q^3 + \sigma q)(\kappa q^4 + \sigma q^2)} \right\} \nonumber \\
   &= \frac{k_B T}{4 \pi\sigma }
\ln\!\left(\frac{\pi^2 \kappa + A\sigma}
{\pi^2 \kappa + b^2\sigma}\right)
+ \frac{\Gamma^a}{4 \pi A \sigma^3}
\left[\frac{(A-b^2)\sigma}{\pi^2}
-\frac{\pi^2\kappa^2}{\pi^2\kappa + A\sigma}
+\frac{\pi^2\kappa^2}{\pi^2\kappa + b^2\sigma}
-2\kappa\ln\!\left(
\frac{\pi^2 \kappa + A\sigma}
{\pi^2 \kappa + b^2\sigma}\right)\right].
\end{align}
\begin{figure}[!t]
  \centering
  \includegraphics[width=0.6\textwidth]{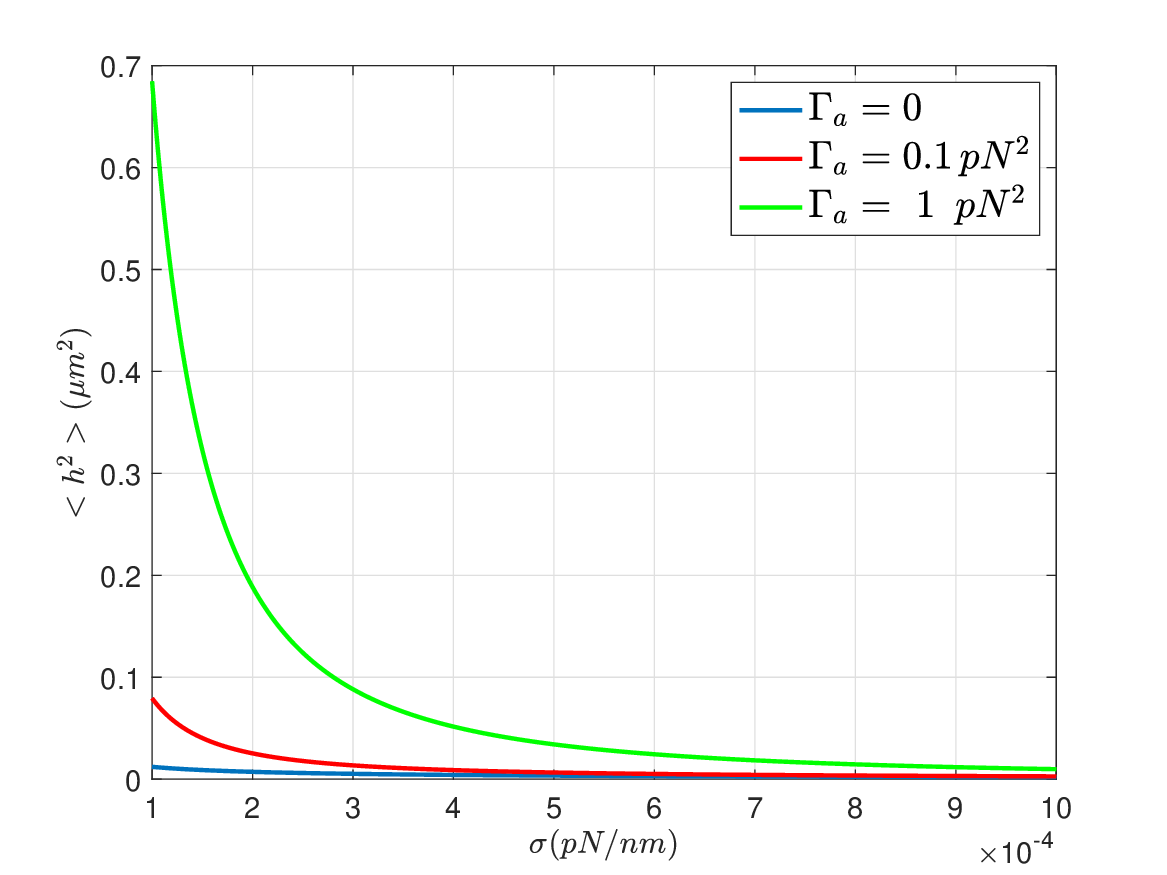}
  \caption{Mean square amplitude of height fluctuation of a membrane against tension for various levels of activity. More activity leads to larger fluctuations.}
  \label{Mean_h2}
\end{figure}
Once again, $\langle h^{2} \rangle$ can be analytically evaluated in terms of $\tan^{-1}$ and $\ln$ functions when $\tau_{a}$ is finite and non-zero, but we omit the expressions here for the sake of brevity. Just as in the tension-area relation, in Eq.~\eqref{eq:finh2} the first term arises due to thermal fluctuations and the second term due to activity. The mean square height fluctuations also scale linearly with $k_{B}T$ and $\Gamma^a$, as did the reduction in area. Figure \ref{Mean_h2} shows the plot for the mean squared amplitude of height fluctuation of a membrane against tension for various levels of activity. We note that more activity leads to a dramatic increase in $\langle h^{2} \rangle$ for the same level of tension consistent with the observation for reduced area in Fig.~\ref{Reduced_area}. 

For a tensionless membrane, with $\sigma = 0$, the mean square amplitude of height fluctuations turns out to be
\begin{equation}
   \langle h^2 \rangle = \frac{A}{2 \pi} \int_{\pi/\sqrt{A}}^{\pi/b} q\, dq\, \left\{\frac{k_B T}{A(\kappa q^4 )} + \frac{ \Gamma^a}{A^2(\kappa q^4)^2} \right\}\\
   = \frac{k_B TA}{4 \pi^3\kappa}
\left(1-\frac{b^2}{A}\right)
+\Gamma_a\,\frac{A^3-b^6}{12A\pi^7\kappa^2}.
\end{equation}
In the absence of activity it is easy to see that the reduction in area is related to the mean square height fluctuations as
\begin{equation} \label{eq:h2sig}
\frac{A_{\text{red}}}{A} = \frac{k_{B}T}{8\pi\kappa}\ln\frac{A}{b^{2}} - \frac{\sigma}{2\kappa}\langle h^{2} \rangle.
\end{equation}
Here, we used Eqs.~\eqref{eq:fintA} and \eqref{eq:finh2} to eliminate $\Gamma^a$ and write $\frac{A_{\text{red}}}{A}$ in terms of $\langle h^{2} \rangle$. Therefore, one can deduce $\langle h^{2} \rangle$ versus $\sigma$ from a measurement of $A_{\text{red}}/A$ versus $\sigma$, or vice-versa. A similar relation can also be found for the tensionless membrane in the presence of activity assuming $A >> b^{2}$
\begin{equation} \label{eq:h2gam}
    \frac{A_{\text{red}}}{A} \approx \frac{k_{B}T}{8\pi \kappa}\ln\frac{A}{b^{2}} + \frac{3\pi^2}{4A}\langle h^{2} \rangle.
\end{equation}
Note that in Eq.~\eqref{eq:h2sig} and Eq.~\eqref{eq:h2gam}, the first term on the right hand side is identical and the second term is linear in $\langle h^{2} \rangle$. Therefore, a naive observer (who does not know of the presence of activity) of the tensionless active membrane would ascribe the increase in $A_{\text{red}}$ due to activity to an in-plane hydrostatic {\it compression} since the signs of the coefficients of $\langle h^{2} \rangle$ in Eqs.~\eqref{eq:h2sig} and \eqref{eq:h2gam} are opposite. 

\section{Correlation of the normal vectors}
Another way to describe fluctuations of membranes is through the scalar product of the normals at different points on the surface. For nearby positions, $\mathbf{n}(\mathbf{r}_1)$ and
$\mathbf{n}(\mathbf{r}_2)$ are approximately parallel, and their dot product is
close to unity. In contrast, for a wavy or fluctuating surface, the normals vary with position, and this
scalar product depends on the separation between points. As the separation increases, the normals become increasingly uncorrelated, and the average dot product decreases toward zero. 

For a fluctuating membrane, the average $\mathbf{n}(\mathbf{r}_1)\cdot \mathbf{n}(\mathbf{r}_2)$ over all configurations can be calculated as \cite{boal2012mechanics},
\begin{equation}
    \langle \mathbf{n}(\Delta \mathbf{r}) \cdot \mathbf{n}(0) \rangle
= 1 - \frac{A}{4\pi^{2}} \int q^{2}\, \mathrm{d}\mathbf{q}\,
\left[ 1 - \cos(\mathbf{q}\cdot\mathbf{r}) \right]
\langle |h_{\mathbf{q}}|^2 \rangle,
\end{equation}
where we integrate the expression with the limits of $q$ ranging from $\pi / \Delta r$ to $\pi / b$ where $\Delta r = |\Delta \mathbf{r}|$. This is equivalent to a two-dimensional integral with $d\mathbf{q} = q\,dq\,d\theta$, (where
$\theta$ is the angle between $\mathbf{q}$ and $\mathbf{r}$, i.e.,
$\mathbf{q}\cdot\mathbf{r} = q|\mathbf{r}|\cos\theta$).
Defining the $\theta$ part of the integral to be $I_\theta(z)$, we have
\begin{equation}
\langle \mathbf{n}(\Delta\mathbf{r}) \cdot \mathbf{n}(0) \rangle
= 1 - \frac{A}{2\pi}
\int q^3 dq \, I_\theta(z) \langle |h_{\mathbf{q}}|^2 \rangle,
\end{equation}
with $z = q|\mathbf{r}|$ and
\begin{equation}
I_\theta(z)
= 1 - \frac{1}{2\pi} \int_{0}^{2\pi} d\theta \, \cos\,\bigl(z\cos\theta\bigr),
\end{equation}
is a Bessel function. At large $z$, this integral can be approximately truncated to $1$, which is applicable here since our limits for $q$ are large. Substituting the expression for $\langle |h_{\mathbf{q}}|^2 \rangle$ for a passive membrane and considering $\sigma = 0$, the integral ultimately reduces to \cite{boal2012mechanics}
\begin{equation}\label{corr-normal-passive}
\langle \mathbf{n}(\Delta \mathbf{r}) \cdot \mathbf{n}(0) \rangle
\sim 1 - \frac{k_B T}{2\pi \kappa}
\int_{\pi/\Delta r}^{\pi/b} \frac{\mathrm{d}q}{q}
= 1 - \frac{k_B T}{2\pi \kappa}
\ln\!\left(\frac{\Delta r}{b}\right).
\end{equation}
We can now evaluate the same expression for an active membrane. Taking the limit $\tau_a \to \infty$ and setting $\sigma = 0$ we get
\begin{eqnarray}\label{corr-normal-active}
\langle \mathbf{n}(\Delta \mathbf{r}) \cdot \mathbf{n}(0) \rangle
&\sim& 1 - \frac{1}{2\pi}
\int_{\pi/\Delta r}^{\pi/b} \left(\frac{k_B T}{\kappa^2 q} + \frac{ \Gamma^a}{A \,\kappa q^5} \right)\mathrm{d}q \nonumber \\
 &\sim& 
 1 - \left(\frac{k_B T}{2\pi \kappa}
 \ln\!\left(\frac{\Delta r}{b} \right) + \frac{\Gamma^a(\left( \Delta r^4\right) -b^4)}
     {8 \pi^5  \kappa^2 A}
\right).
\end{eqnarray}
When $\Delta r = b$ the correlation $\langle \mathbf{n}(\mathbf{\Delta r})\cdot \mathbf{n}(0)\rangle = 1$; this is expected since $b$ is typically the size of a lipid headgroup over which the normal cannot change. Equation~\eqref{corr-normal-active} shows that for $\Delta r > b$, there is a larger decrease in the correlation of normals when $\Gamma^{a} > 0$. Said differently, the presence of activity causes a faster decay in the correlation of normals, as expected. This is because active noise increases the fluctuations of the active membrane.

\begin{figure}[!htbp]
  \centering
  \includegraphics[width=0.6\textwidth]{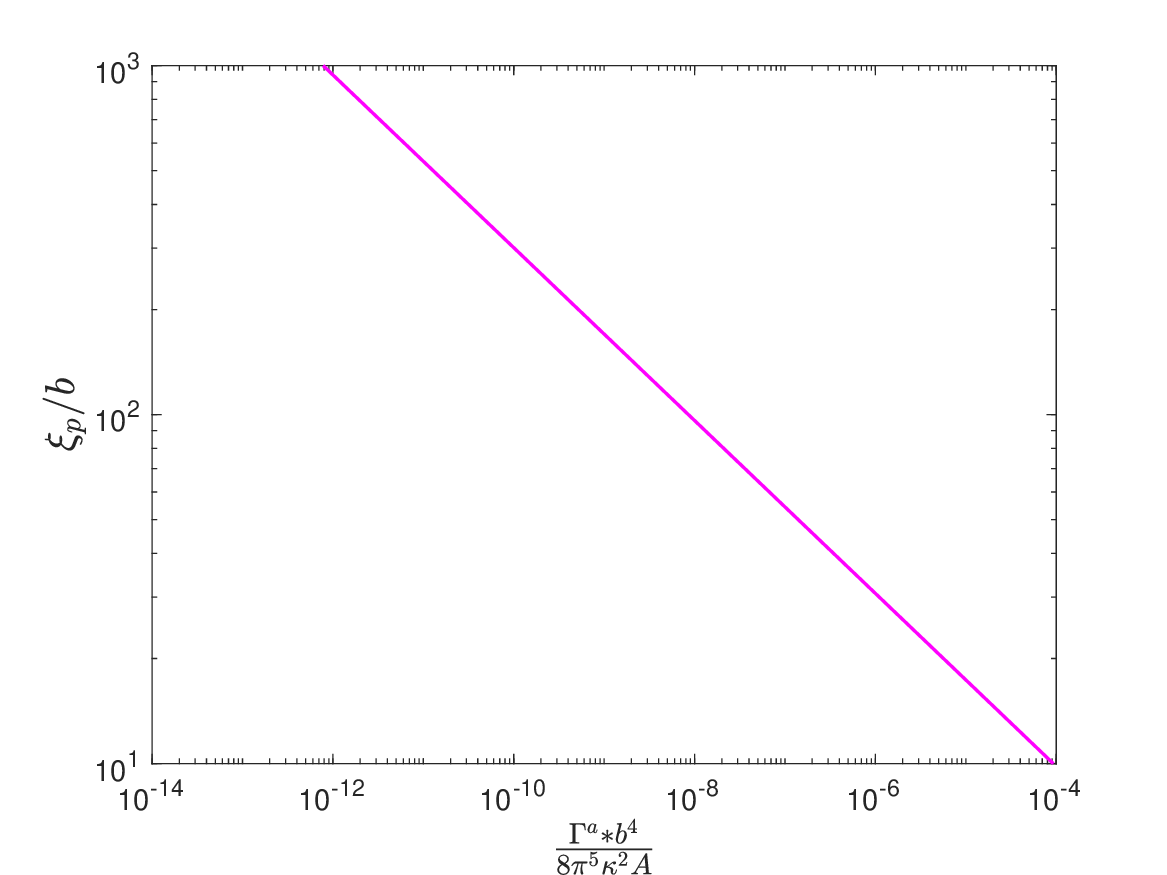}
  \caption{A plot for persistence length against normalized activity. As activity increases the persistence length decreases because the increased fluctuations causes the memory of normals to be forgotten over a shorter distance. }\label{Persistence_length_normalized}
\end{figure}
\begin{figure}[!htbp]
  \centering
  \includegraphics[width=0.6\textwidth]{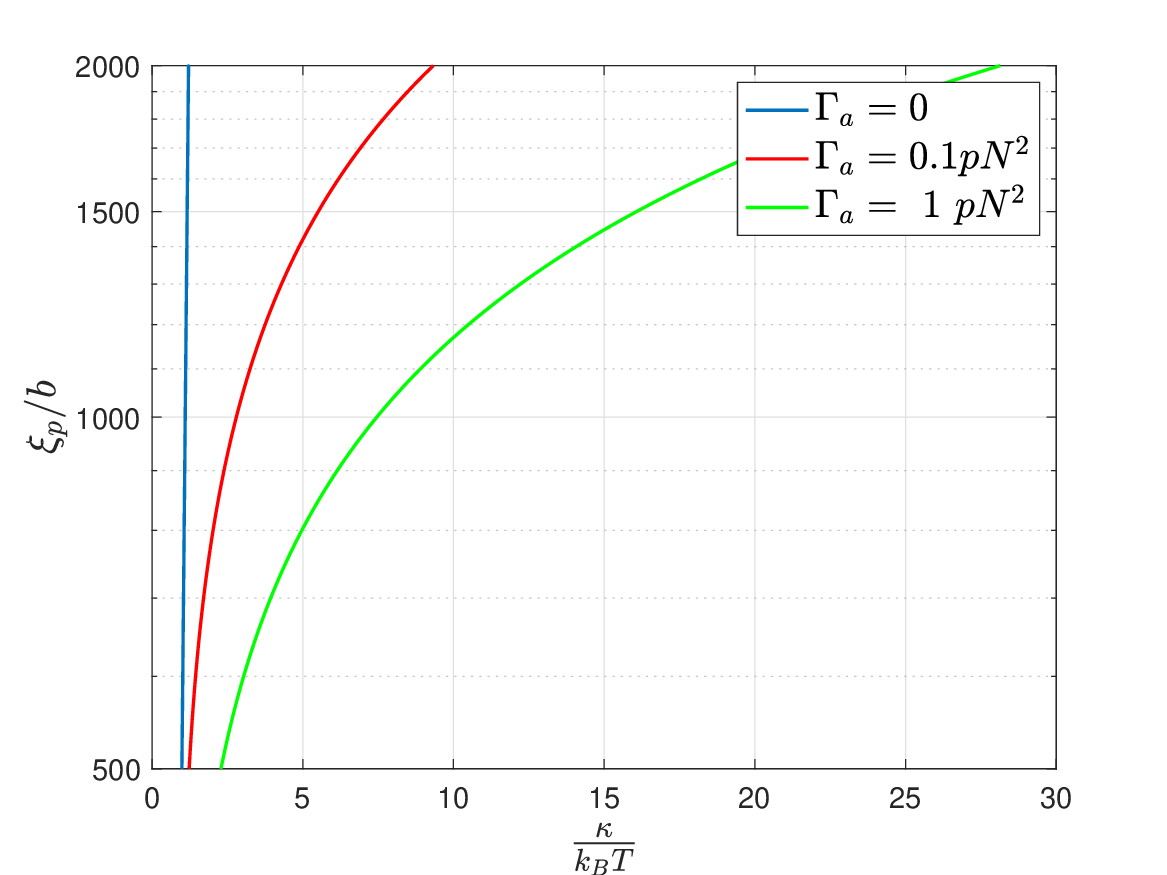}
  \caption{A plot for persistence length against normalized bending modulus. An increased bending modulus results in longer persistence length of the membrane, just as in filaments. Increased activity slows the rate of increase of persistence length as bending modulus increases. }\label{fig:persvskap}
\end{figure}

\section{Persistence Length}
Persistence length defines the length scale over which a filament or membrane remains ``stiff'' before thermal fluctuations cause it to lose its directional memory \cite{boal2012mechanics, Phillips2013}. In the context of filaments, this is the distance at which the two ends of the filament have uncorrelated orientations as measured by the tangent vectors. In the context of membranes, the quantity naturally analogous to the correlation of the tangent vectors in filaments is the correlation of the surface normal vectors. Hence, we use the expression derived in the previous section to estimate the persistence length of an active membrane. 

Following \cite{boal2012mechanics}, the persistence length, $\xi_{p}$, is defined as the distance $\Delta r$ at which $\langle \mathbf{n}(\Delta r)\cdot\mathbf{n}(0)\rangle$ given by Eqs. \eqref{corr-normal-passive}  and \eqref{corr-normal-active} decays to zero. We solve for $\xi_{p}$ by setting the left hand side of Eq.~\eqref{corr-normal-active} to zero: 
\begin{equation}
1 - \left(\frac{k_B T}{2\pi \kappa}
 \ln\!\left(\frac{\xi_{p}}{b} \right) + \frac{\Gamma^a(\xi_{p}^4 -b^4)}
     {8 \pi^5  \kappa^2 A}\right) = 0.
\end{equation}   
Figure \ref{Persistence_length_normalized} shows $\xi_{p}/b$ as a function of normalized $\Gamma^a$. We find that as the activity increases the persistence length of the membrane decreases, again as expected, because increased fluctuations obliterate directional memory of surface normals. In Fig.~\ref{fig:persvskap} we plot the normalized persistence length as a function of bending modulus for various levels of activity. Persistence length increases as the bending modulus increases, but its rate of increase decreases as the activity increases.

\section{Conclusions}

In this work, we presented a non-equilibrium statistical mechanics framework to investigate the mechanical behavior of active membranes. By utilizing a variational approach combined with an over-damped Langevin equation, we derived analytical expressions that proffer insights into four fundamental physical metrics that characterize the membrane's mechanical response. 
Specifically, the derived tension–area relation reveals that activity significantly enhances the reduction in projected area, as energy-consuming processes drive more of the membrane into out-of-plane fluctuations. This is further supported by our analysis of the mean square amplitude of fluctuations, which shows a linear scaling with active force strength and reveals a unique relationship with the projected area that could lead to an interpretation of the active noise as hydrostatic compression.
We also found that active noise leads to a faster spatial decay in the correlation of surface normals compared to passive membranes. Consistent with the findings on normal vector correlations, we established that increased activity directly reduces the persistence length of the membrane. This quantifies how active processes effectively ``soften'' the membrane's long-range structural integrity. 
Collectively, these analytical expressions bridge the gap between microscopic active processes and macroscopic mechanical observables. Because active and thermal fluctuations often result in similar linear scaling of observables, our study provides a theoretical foundation for interpreting experimental data from living matter and distinguishing the signatures of life or activity from those of simple thermal equilibrium.

\section*{Acknowledgment} 

We acknowledge the support from the U.S.\ National Science Foundation, Division of Civil, Mechanical, and Manufacturing Innovation (Award No.\ 2227556 to YK), and Division of Materials Research (Award No.\ 2212162 to PKP).

\bibliography{StatMech_bib}

\end{document}